\documentclass[aps,preprint,amssymb,12pt,floatfix]{revtex4}
\usepackage{subfigure}
\usepackage{graphicx}
\usepackage{rotating}
\usepackage{color}

\begin{document}
\title{Mechanical unfolding of RNA: From hairpins to structures with internal multiloops}
\author{Changbong Hyeon$^1$ and D. Thirumalai$^{1,2}$}
\thanks{Corresponding author phone: 301-405-4803; fax: 301-314-9404; 
thirum@glue.umd.edu}
\affiliation{$^1$Biophysics Program\\
Institute for Physical Science and Technology\\
$^2$Department of Chemistry and Biochemistry\\
University of Maryland, College Park, MD 20742\\
}

\date{\today}
\baselineskip = 14pt

\begin{abstract}
Mechanical unfolding of RNA structures, ranging from hairpins to ribozymes, 
using laser optical tweezer (LOT) experiments have begun to reveal the features of the energy landscape that cannot be easily explored using conventional experiments.  
Upon application of constant force ($f$), RNA hairpins undergo cooperative transitions from folded to unfolded states  
whereas subdomains of  ribozymes unravel one at a time.  
Here, we use a self-organized polymer (SOP) model and Brownian dynamics simulations to probe mechanical unfolding at  constant force and  
constant-loading rate  of four RNA structures of varying complexity. 
 For simple hairpins, such as P5GA,  application of constant force or constant loading rate  results in bistable cooperative transitions between folded and unfolded states 
without populating any intermediates.  
The transition state location ($\Delta x_F^{TS}$) changes dramatically as the loading rate is varied.  
At loading rates that are comparable to that used in LOT experiments the hairpin is plastic with $\Delta x_F^{TS}$ being midway between folded  
and unfolded states whereas at high loading rates $\Delta x_F^{TS}$ moves close to the folded state i.e, RNA is brittle. 
For the 29 nucleotide TAR RNA with the three nucleotide bulge unfolding occurs in a nearly two state manner with an occasional pause 
in a high free energy metastable state.  
Forced unfolding of the 55 nucleotide of the Hepatitis IRES domain IIa, which has a distorted L-shaped structure, results in well populated stable intermediates. 
The most stable force-stabilized intermediate represents straightening of the L-shaped structure.  
For  these structures the unfolding pathways can be predicted using the contact map of the native structures.  
Unfolding of a RNA motif with internal multiloop, namely, the 109 nucleotide prohead RNA which is part of the $\phi$29 DNA packaging motor,
at constant value of $r_f$ occurs with three distinct rips that represent unraveling of the paired helices.  The rips represent kinetic barriers to unfolding.
Our work shows (i) the response of RNA to force
is largely determined by the native structure; (ii) only by probing mechanical unfolding over a wide range of forces can the underlying
energy landscape be fully explored.  
\end{abstract}
\maketitle

\newpage

\section{INTRODUCTION}

The discovery of self-splicing catalytic activity of ciliate \emph{Tetrahymena 
thermophila} 
ribozyme \cite{CechCell81,AltmanSci84} and subsequent findings that RNA 
molecules play 
an active role as enzymes \cite{DoudnaNature02} 
in many cellular processes have revolutionized RNA research. 
Just as for protein folding the structures and functions of RNA enzymes (ribozymes) are linked.
As a result, the RNA folding problem, namely, how a nucleotide sequence folds to 
the native state conformation 
is important in molecular biology. 
Several studies from a number of groups 
\cite{TreiberCOSB01,SosnickCOSB03,PanJMB97,HyeonBC05,ZhuangACR05,HerschlagJMB99} have shown 
 that, under \emph{in vitro} conditions, ribozymes 
have rugged energy landscapes.  
Despite significant advances in our understanding of how ribozymes fold, 
several outstanding issues remain 
which can be addressed using single molecule experiments 
\cite{TreiberCOSB01,SosnickCOSB03,PanJMB97,HyeonBC05,ZhuangACR05,HerschlagJMB99,ZhuangSCI00}. 

Thermodynamic and kinetic measurements in ensemble and single molecule florescent energy transfer experiments are typically 
made by varying the concentration of counterions. 
Recently, using laser optical tweezer setup mechanical force has been used to 
trigger folding and unfolding of RNA molecules at a single 
molecule level \cite{Bustamante2,Bustamante4}. 
Mechanical force, applied to a specific position of the molecule, induces  
sequence and structure-dependent response, 
which is reflected in the force-extension curve (FEC) that is usually fit using the worm-like (WLC) model \cite{BustamanteSCI94,MarkoMacro96}. 
The stability of RNAs is inferred by integrating the FECs. 
For simple motifs, such as hairpins, it has been shown that the stability of the 
native structures can be 
accurately measured using mechanical unfolding trajectories which exhibit multiple 
transitions between the folded and the unfolded state when the force is held constant \cite{Bustamante2}.  
Similarly, thermodynamics of ribozymes can also be obtained using the 
non-equilibrium work theorem \cite{JarzynskiPRL97,Bustamante3}. 

Mechanical force has also been used to probe unfolding and refolding kinetics of 
RNA. 
The cooperative reversible folding of hairpins has been shown by monitoring the 
end-to-end distance ($R$), 
a variable conjugate to the mechanical $f$, as a function of time. 
This procedure works best when RNA folding is described using two-state 
approximation.
For multidomain ribozymes the folding/unfolding kinetics is complex and new 
tools are 
required to interpret the kinetic data. 
In a pioneering study Onoa \emph{et. al.} \cite{Bustamante4} showed that the 
rips in FECs for the L-21 derivative 
of \emph{Tetrahymena thermophila} ribozyme (\emph{T.} ribozyme),
composed of multiple domains, are a result of unfolding of individual intact 
domains 
that are stabilized in the native state by counterion-dependent tertiary interactions. 

The single molecule studies show that the response to mechanical force is a 
powerful tool 
to analyze the underlying principles of RNA self-assembly. 
The extraction of unfolding pathway using FECs alone is not  easy  
especially when 
the ribozyme is composed of multiple domains \cite{Bustamante4}.
To decipher the unfolding pathways  of \emph{T}. ribozyme Onoa 
\emph{et. al.} \cite{Bustamante4}  
did a series of experiments in which FECs of different independently-folding subdomains were used to 
interpret the order of 
unfolding of the substructures. 
In RNA, there is a clear separation in the free energies associated with 
secondary and tertiary interactions. 
Thus, the FEC for a multi-domain ribozyme is, to a first approximation, the 
union of the FECs for the individual domains. 
Such a strategy can be used to assign a rip of FEC to the unfolding of a 
particular subdomain 
as long as the contour lengths of two different unfolded motifs are not similar.  
Moreover, it is known that the precise response of RNA to force depends not only 
on the sequence and the native structure 
but also on how the force is applied \cite{HyeonPNAS05,TinocoBJ06}. 
Single molecule experiments can be performed in different modes that includes either 
force-clamp ($f$ is in constant) \cite{FernandezSCI04,Bustamante2} or force-ramp ($f$ varies in a time dependent manner) 
\cite{TinocoBJ06,FernandezPNAS99}. 
Theoretical studies have proposed models for obtaining a number of experimentally measurable quantities including FECs for RNA 
\cite{HwaBP03,HwaBP01,MarkoEPJE03}. 
Computational studies have shown, using RNA hairpin as an example, that the 
kinetics of unfolding and force-quench refolding as well the nature of unfolding Ädepend 
on the magnitude of $f$ and the loading rate ($r_f$) 
\cite{HyeonPNAS05,HyeonBJ06}. 
These studies show that it is important to complement the single molecule studies with 
computations that can reliably resolve key 
issues that are difficult to address in experiments. 

In this paper, we probe the forced-unfolding dynamics of RNA molecules using 
a simple model. 
Because these simulations can be used to directly monitor structures in the transition from folded to the fully stretched states, unfolding pathways 
can be unambiguously resolved.  
We introduce the \emph{Self-Organized Polymer (SOP) model} for RNA that is based 
only 
on the self-avoiding nature of the RNA and the native structure. 
We apply the SOP model to probe forced-unfolding of a number of RNA structures of varying complexity.
Many of the 
subtle features of the variations in the mechanical unfolding as a function of 
$f$ and $r_f$ 
can be illustrated using P5GA, a simple RNA. 
For example, we show that the dramatic movements in the location of the 
unfolding transition state occur as $r_f$ (or $f$) is varied. 
Applications to structures of increasing complexity (TAR RNA, prohead 
RNA from domain IIa of the Hapatitus C virus, 
$\phi$29 DNA bacteriophage motor) show that 
discrete intermediates can be populated in force-ramp and force-clamp 
simulations over a certain range of forces.  Our results  show that the response
of RNA to force is largely dependent on the architecture of the native state.  More importantly,
we have established that the characterization of the the energy landscape requires using force values
(or loading rates) over a wide range. \\

\section{METHODS}

{\it Model:} 
Our goal is to construct a model for obtaining mechanical folding and unfolding 
trajectories for simple RNA hairpins to 
large ribozymes. 
The model has to be realistic enough to take into account the interactions that 
stabilize the native fold yet simple 
enough so that the response to a wide range of forces and loading rates can be 
explored. 
To this end, we introduce a new class of versatile coarse-grained self-organized 
polymer (SOP) model that is 
particularly well suited for single-molecule force spectroscopy applications of 
large ribozymes and 
proteins. 
The SOP model can be used to probe the response of mechanical force that is 
applied by means of 
force-clamp (constant force), force-ramp, and force-quench. 
The reasons for using SOP model in force spectroscopy applications are the 
following :
(1) Forced-unfolding, and force-quench refolding lead to large conformational 
changes. 
For example, upon application of constant force the end-to-end distance of the 
RNA changes by about 
(10-100) $nm$ 
depending on the size of RNA. 
Currently, single molecule experiments (LOT or AFM) cannot resolve structural 
changes below a few $nm$. 
As a result, details of the rupture of hydrogen bonds or 
local tertiary contacts between specific bases cannot be discerned 
from FEC or the dynamics of $R$ alone. 
Because only large changes in $R$, the variable that is conjugate to force, are 
monitored it is not 
crucial to include details of the local interactions such as bond-angle and 
various dihedral angle potentials. 
(2) We had shown, in the context of mechanical unfolding of proteins, that many 
of the details of unfolding pathways can be accurately 
computed by taking into account interactions that stabilize the native fold 
\cite{Klimov2}. 
Based on this observation accurate predictions of unfolding forces and the 
location of the unfolding transition states were made for a number of proteins 
with $\beta$-sandwich, $\alpha/\beta$, and $\alpha$-helical folds. 
Our previous study \cite{Klimov2} also suggested that it is crucial to take into account chain 
connectivity and attractive interactions that 
faithfully reproduce the contact map of a  fold. 
(3) Electrostatic interactions are pivotal in RNA. 
However, under physiological condition, counterion concentration is large enough 
to 
effectively screen the electrostatic repulsion between the phosphate groups. 
Thus, due to effective screening (small Debye length) the repulsive 
electrostatic potential between phosphate groups is 
effectively short ranged. 

With the above observations in mind we propose, the SOP model for RNA that 
retains chain connectivity and 
favorable attractive interactions between sites that 
stabilize the native fold. 
Each interaction center represents the center of mass of a nucleotide. 
In terms of the coordinates $\{\textbf{r}_i,i=1,2,\ldots N\}$ of RNA with $N$ 
nucleotide the total potential energy in the SOP representation is 
\begin{eqnarray}
V_{T}&=&V_{FENE}+V_{nb}^{(att)}+V_{nb}^{(rep)}\nonumber\\
&=&-\sum_{i=1}^{N-1}\frac{k}{2}R_0^2\log({1-\frac{(r_{i,i+1}-r_{i,i+1}^o)^2}{R_0
^2}})\nonumber\\
&+&\sum_{i=1}^{N-3}\sum_{j=i+3}^N\epsilon_h[(\frac{r^o_{ij}}{r_{ij}})^{12}-2(\frac{r^o_{ij}}{r_{ij}})^6]\Delta_{ij}\nonumber\\
&+&\sum_{i=1}^{N-2}\epsilon_l(\frac{\sigma^*}{r_{i,i+2}})^6+\sum_{i=1}^{N-3}\sum_{j=i+3}^N\epsilon_l(\frac{\sigma}{r_{ij}})^6(1-\Delta_{ij}).
\label{eq:SOP}
\end{eqnarray}
The first term is for the chain connectivity. 
The finite extensible nonlinear elastic (FENE) potential \cite{KremerJCP90} is 
used 
with $k=20kcal/(mol\cdot$\AA$^2)$, $R_0=0.2$ $nm$, 
and $r_{i,i+1}$ is the distance between neighboring beads interaction centers 
$i$ and $i+1$, $r^o_{i,i+1}$ is the 
distance in the native structure.
The use of FENE potential is more advantageous than the standard harmonic 
potential especially when considering forced-stretching 
because the fluctuations of $r_{i,i+1}$ are strictly restricted 
around $r_{i,i+1}^o$ with variation of $\pm R_0$.   
The Lennard-Jones potential is used to account for interactions that stabilize 
the native topology. 
Native contact is defined for the pair of interaction centers whose distance is 
less than $R_C=1.4$ $nm$ in the native state for $|i-j|>2$. 
If $i$ and $j$ sites are in contact in the native state, $\Delta_{ij}=1$, otherwise 
$\Delta_{ij}=0$.
We used $\epsilon_h=0.7$ $kcal/mol$ for the native pairs, 
$\epsilon_l=1$ $kcal/mol$ for non-native pairs.  In the current version, we have neglected non-native interactions which will not
qualitatively affect the results because, under tension, such interactions are greatly destabilized. 
To ensure the non-crossing of the chain, we set $\sigma=7$\AA. 
Only for $i,i+2$ pairs we set $\sigma^*=3.5$ \AA\ to prevent the flattening of the 
helical structures when the overall repulsion is large. 
There are five parameters in the SOP force field ($k$, $R_0$, $\epsilon_h$, $\epsilon_l$, and $R_c$) \cite{HyeonSTRUCTURE06}.
Of these the results are sensitive to the precise values of 
$\epsilon_h/\epsilon_l$ and $R_c$. 
We have discovered that the quantitative results are insensitive to 
$R_c$ as long as it is in the physical range that is determined by the RNA contact 
maps. 
In principle, the ratio $\epsilon_h/\epsilon_l$ can be adjusted to obtain 
realistic values of forces. 
For simplicity we choose a uniform value of  $\epsilon_h$  for all RNA constructs.  Surprisingly, the SOP force field, with the same set of parameters, can be used to
obtain near quantitative results for RNA molecules of varying native topology.

The time spent to calculate Lennard-Jone forces scales as $\mathcal{O}(N^2)$.
Drastic savings in computational time can be achieved by truncating the forces 
due to the 
Lennard-Jones potential for interaction pairs with $r_{ij}>(3r_{ij}^o$ or 
$3\sigma)$, 
to zero. 
We refer to the model as ``Self-Organized Polymer (SOP) model'' because it only 
uses the polymeric nature of the biomolecules 
with the crucial topological constraints that arise by the specific fold. 
For probing forced-unfolding of RNA (or proteins) it is sufficient to include attractive interactions only 
between contacts that stabilize the native state (see Eq. \ref{eq:SOP}). 
We believe none of the results will change qualitatively if this restriction is relaxed i.e., if non-native interactions are also taken into account. 

{\it Simulations:} 
Using the SOP model, we simulated the mechanical unfolding and refolding of 
various RNA structures from a 
simple hairpin to a large ribozyme ($N\approx 400$). 
To simulate force-ramp experiments we pull a harmonic spring ($k_s=28$ $pN/nm$) 
that is attached 
to the 3' end of molecule at a constant speed ($v$).
The time($t$)-dependent  force acting on the 3' end is $f=-k_s(z-vt)$ where $z$ is $z^{th}$ 
coordinate of the 3' end. 
In force-clamp simulations a constant force is applied to one end of the 
molecule while the other end is fixed. 
Finally, in force-quench computations the force on the molecule is reduced to 
the final value to initiate mechanical refolding. 
In both force-clamp and force-quench setup the dynamics of the linker (usually 
hybrid RNA/DNA handles) is not relevant whereas 
depending on the characteristics of the linkers the dynamics of linker may play 
an important role in the 
force-ramp experiments \cite{HyeonBJ06}. 

{\it Time scales:}
Since a typical value for the mass of a nucleotide,
$m\sim 300g/mol-400g/mol$,
the average distance between the adjacent nucleotides in the SOP representation 
of RNA is 
$a\approx 5$ \AA, $\epsilon_h=0.7$ $kcal/mol$, 
the natural time is $\tau_L=(\frac{ma^2}{\epsilon_h})^{1/2}=3\sim5$ $ps$.
We use $\tau_L=4.0ps$ to convert simulation times into real times.
To estimate the time scale for mechanical unfolding dynamics 
we use a Brownian dynamics algorithm \cite{McCammonJCP78,VeitshansFoldDes96}
for which the natural time for the overdamped motion is 
$\tau_H=\frac{\zeta\epsilon_h}{k_BT}h\tau_L$.
We used $\zeta=100\tau_L^{-1}$ in the overdamped limit, that approximately 
corresponds to friction constant of 
a nucleotide in water. 

The equations of motion in the overdamped limit are integrated using the 
Brownian dynamics algorithm. 
The position of a bead $i$ at the time $t+h$ is given by 
\begin{equation}
x_i(t+h)=x_i(t)+\frac{h}{\zeta}(F_i(t)+\Gamma_i(t))
\end{equation}
where $F_c(t)=-\frac{\partial V}{\partial x}$, the Newtonian force acting on a 
bead $i$, 
and $\Gamma(t)$ is a random force on $i$-th bead that has a white noise 
spectrum. 
The autocorrelation function for $\Gamma(t)$ in the discretized form is 
\begin{equation}
\langle\Gamma_i(t)\Gamma_j(t+nh)\rangle=\frac{2\zeta 
k_BT}{h}\delta_{0,n}\delta_{i,j}
\end{equation}
where $\delta_{0,n}$ is the Kronecker delta function and $n=0,1,2,\ldots$. 
All the force simulations are performed at $T=300$ $K$.
For the integration time step $h=0.1\tau_L$, $10^6$ integration time steps in the over-damped limit ($\zeta = 100 \tau_L^{-1}$) is 
$10^6 \tau_H = 10^6 \frac{\zeta \epsilon_h}{k_BT} h \tau_L =47 \mu s$ with $\tau_L = 4ps$, $\epsilon_h \approx 0.7kcal/mol$, and $k_BT \approx 0.6 kcal/mol$.
The system composed of RNA and the spring is extended along the force direction 
by 
$\delta x$  every $10^4 \tau_H= 0.47 \mu s$ integration time steps. 
We chose $\delta x=0.003 nm$ so that the pulling speed,
$v= \frac{0.003nm}{0.47 \mu s} =6.4$ $\mu m/s$ for $h=0.1\tau_L$.  
In order to maintain numerical stability neither $h$ nor $\delta x$ 
should be too large.

{\it Contact map:} 
In RNA with simple native structures force-induced unfolding pathways can be 
qualitatively predicted 
from the native structure. 
To rationalize the simulated unfolding pathways it is useful to construct RNA 
contact maps. 
We generated the contact maps 
(Figs. \ref{P5GA_cf}, \ref{1uud_analysis}, \ref{1p5m_analysis}) 
using the definition of native contact. 
More precisely, we generated a matrix $\mathcal{Q}$ where matrix elements are 
\begin{equation}
\mathcal{Q}_{ij}=\Theta(R_c-r_{ij}^o)
\end{equation}
where $r_{ij}^o$ is the distance between two nucleotides in the SOP 
representation 
of the native fold. 
Representing the native RNA using only the center of mass of each nucleotide as 
the interaction center 
is a drastic simplification. 
To ascertain if the SOP representation misses any essential feature of the RNA 
structure we also generated
the distance map using the heavy atom (C, N, O, P) coordinates. 
The coarse-grained model captures the important interactions 
on length scales that is larger than about 0.7 $nm$. 
For example, the SOP contact map (Fig.\ref{1uud_analysis}-C) 
and the distance map (Fig.\ref{distancemap_1uud}) of $1uud$ are similar. 

{\it Dynamics of rupture of contacts:}
The dynamics of RNA unfolding is monitored using a number of variables including 
the time dependence of $R$ and the number of nucleotide-dependent native 
contacts $Q_i(t)$ that remain at time $t$. 
We define $Q_i(t)=\sum_{j(|j-i|>2)}^N\Theta (R_C-r_{ij}(t))\Delta_{ij}$ where 
$R_C$ is the cut-off distance for native contacts, $r_{ij}(t)$ is
the distance between $i$ and $j$-th nucleotide, and  $\Delta_{ij}=1$ for native 
contact otherwise $\Delta_{ij}=0$.
If a certain subdomain of the molecule is disrupted and loses its contacts,
then the extension of the molecule suddenly increases and the mechanical force 
exerted on the end of the molecule drops instantly.
These molecular events are reflected as rips in the FEC.
When the time dependence of the force $f(t)$ or the end-to-end distance $R(t)$ 
is directly compared with $Q_i(t)$ using $t$ as a progressive variable to 
describe unfolding,
the direct correlation between sudden drops (sudden increase) in the value of 
$f(t)$ ($R(t)$) and $Q_i(t)$ enables us to unambiguously identify the structures 
involved in the 
dynamics of rupture of contacts at the nucleotide level.

\section{RESULTS AND DISCUSSION}

{\bf Mechanical unfolding of the secondary structural elements of RNA}

The stability of RNA molecule in the native state can be approximated 
as the sum of interactions 
$\epsilon^{tot}=\sum_i\epsilon^{sec}_i+\sum_k\epsilon^{ter}_k$ 
where $\epsilon^{sec}_i$ and $\epsilon_k^{ter}$ are interactions that stabilize 
the secondary and tertiary structures, respectively, $i$ refers to the number of 
secondary structural elements, and $k$ labels the tertiary contacts that 
may be mediated by counterions.  
The contributions from the tertiary interactions are 
small compared to the energetics associated with the secondary interactions 
($\sum_i\epsilon_i^{sec}\gg\sum_k\epsilon_k^{ter}$). 
Because of the stability gap between the secondary and the tertiary interactions 
the analysis of FEC for RNA can be independently made domain by domain. 
The hairpin stacks, that can vary in the length and sequence, are among the 
simplest structural motifs.
Additional structural complexity in RNA arises due to the presence of hairpin 
loops, bulges, 
internal loops, internal multi-loops. 
The remarkable structural diversity of RNA secondary structures allows us to 
probe the sequence 
and fold-dependent energy landscape using force as a perturbation. 
Here, we discuss the force spectroscopy of relatively simple RNA motifs using four examples. 
Many aspects of the physics of mechanical unfolding of RNA, such as the shifts 
in the transition state locations as $r_f$ is changed, can be understood using 
these simple structural motifs as examples. 
\\

\textbf{Force-induced transitions in a simple hairpin (P5GA):} 
Liphart \emph{et. al.} showed that the P5ab hairpin, the construct in which P5c 
stem-loop and the A-rich bulge in P5a 
are removed from the P5abc subdomain in $T$. ribozyme, reversibly folds in an 
all-or-none fashion upon application of constant force. 
The equilibrium between the native basin of attraction  (\textbf{NBA}) and
the unfolded basin of attraction  (\textbf{UBA}) can be shifted by
altering the value of the constant force, $f_c$.
To probe the two-state behavior of hairpins under force we used a smaller 22-nt hairpin, P5GA (Protein Data Bank (PDB) id: 1eor) \cite{TinocoJMB2000,HyeonPNAS05}. 
For the P5GA hairpin simulations over a wide range of forces can be performed in reasonable times.
The topologically simple hairpin has a single tetra-loop and nine consecutive 
base pairs.
In an earlier study \cite{HyeonPNAS05} we showed, using a minimal three 
interaction site
(TIS) model in which each nucleotide is represented by three sites, that
the dynamical behavior of P5GA under tension is qualitatively similar to P5ab.  
The much
simpler SOP representation of P5GA allows us to probe exhaustively the folding 
and
unfolding kinetics of the hairpin that is manipulated by force-ramp,
force-quench, and force-clamp.  

\textit{Constant force:} The hallmark of P5ab \cite{Bustamante2} and P5GA \cite{HyeonPNAS05}, when a
constant force is applied to either the 3' or the 5' ends, is the
observation of bistable kinetics.  When a constant $f_c$ is applied to the
3' end P5GA makes transitions (Fig.\ref{P5GA_cf}-B) between the \textbf{UBA} 
($R\approx 8nm$) to the \textbf{NBA} ($R \approx 2nm$).  At $f_c =$ 14.0, 15.4, 
17.5 $pN$ a large
number of transitions occur over $45 ms$ duration which suggests
that the hairpin dynamics is effectively ergodic.  As in our previous study \cite{HyeonPNAS05} the equilibrium constant between the folded and unfolded hairpin calculated using a long
mechanical unfolding trajectory coincides with an independent ensemble average calculation i.e., time averages are roughly equivalent to ensemble averages.
When $f_c=14$ $pN$ the residence time in the \textbf{NBA} is much greater than in the 
{\bf UBA} 
while at $f_c=16.8$ $pN$ 
the \textbf{UBA} is preferentially populated (Fig. \ref{P5GA_cf}-B). 
The population of P5GA in the \textbf{NBA} changes 
when $f_c$ is varied can be seen in the histogram ($P(R)$) of the
end-to-end distance $R$ (Fig.\ref{P5GA_cf}).  
At $f_c = 15.4$ $pN$, which is slightly above the
midpoint of the \textbf{NBA} $\Leftrightarrow$ \textbf{UBA} transition,
several jumps between the \textbf{NBA} and \textbf{UBA} are observed. 
The $P(R)$ distribution reflects the bistable nature of the
landscape.  
The free energy profile with respect to $R$ is computed using 
$\Delta F(R)=-k_BT\log{P(R)}$.  
From $P(R)$ at $f_c = 15.4$ we can obtain the free energy of stability of the folded hairpin with respect to the unfolded state using $\Delta G \approx f_c \Delta R_{UF}$ 
where $\Delta R_{UF}$ is the distance between the folded and unfolded states of P5GA. 
Using $\Delta R_{UF} \approx 6nm$ we find that $\Delta G \approx 13kcal/mol$.  
The Vienna RNA package \cite{HofackerNAR03}, which uses an entirely different free energy parameters for RNA, gives $\Delta G \approx 12.8 kcal/mol$. 
This comparison shows that the SOP model can, for simple structures, give accurate results for stability.  At $f_c=15.4$ $pN$, the transition barrier is 
about $\sim 1.5 k_BT$. 
The \textbf{UBA} is more populated at this value of $f_c$.
The observed transition times are much shorter than the
residence times in each basin of attraction which is also a reflection of the
underlying cooperativity of the all-or-none of nature of hopping  between {\bf UBA} and 
{\bf NBA}. 

Qualitatively similar results were observed in our previous study using the three interaction site (TIS) 
model \cite{HyeonPNAS05}.
However, the values of the midpoint of the force was about a factor of two
smaller in the TIS model than in the SOP model.  Despite the large
differences in the nature of the force fields the
overall results are robust which suggests that it is the underlying native
structure that determines the nature of the force-induced transitions in simple 
RNA. 
Indeed, for simple structures the mechanism of forced unfolding in RNA helices 
are imprinted in the contact map (Fig.\ref{P5GA_cf}-A) 
which is a two dimensional representation of the folded hairpin (Methods). 
The clustered band in the contact map suggests that P5GA should unfold when a 
critical number of stacking 
interactions are unzipped. 
Thus, the contact map for P5GA (and presumably P5ab) is consistent with the 
observed two-state kinetics. 
As the architecture of the native state becomes more complex it becomes 
difficult to anticipate the unfolding mechanism using the contact map alone (see 
below). 

\textit{Force-ramp:} We also performed force-ramp simulations by subjecting the P5GA hairpin to
a continuously changing force, i.e., 
varying the loading rate (Methods). 
The simplicity of the SOP model allows us to use
values of $r_f$ that are comparable to those  used in LOT experiments. 
At $r_f=45$ $pN/s$ ($\sim 10r_f^{LOT}$) the force-extension curves 
show a transition to the {\bf UBA} at around $f\sim 13$ $pN$ 
(Fig.\ref{P5GA_rf}-A).  
As the force dynamically increases we observe bistable fluctuations in the FEC 
between the {\bf NBA} and the {\bf UBA} just as when force is held constant
(Fig.\ref{P5GA_rf}-A). 
The conformational fluctuations between the two states are unambiguously seen 
in the time-dependence of the end-to-end distance ($R(t)$) 
(Fig.\ref{P5GA_rf}-B). 
As time progresses the force is ramped up which results in global unfolding 
($R\approx 8$ $nm$) 
for $t>400$ $ms$ (Fig.\ref{P5GA_rf}-B). 
During the time scale of simulation we find frequent and sharp transition 
between the {\bf UBA} and the {\bf NBA} (Fig. \ref{P5GA_rf}-B). 

The location of the unfolding transition state $\Delta x_F^{TS}$ for proteins 
and RNA is often estimated 
from force-ramp experiments using the variation of the most probable rupture 
force with $r_f$ ($[f^*,\log{r_f}]$ plot).
The loading rate, $r_f=df(t)/dt$, and can be accurately estimated from the slope 
of the time dependence 
of $f(t)$ as a function of time. 
The slope of $f(t)$ as a function of $t$ (Fig.\ref{P5GA_rf}-C) is nearly the 
same as  $r_f\approx k_s\times v$, where $v$ is the pulling speed. 
Strictly speaking, $r_f= k_{eff}\times v$ with  
$k_{eff}^{-1}=k_s^{-1}+k_{mol}^{-1}+k_{linker}^{-1}$ and $k_{s}$, $k_{mol}$, and 
$k_{linker}$ are the spring constants 
of the optical trap, the RNA molecule, and linker, respectively. Typically 
$k_s\ll k_{mol}, k_{linker}$, thus $k_{eff}\approx k_s$ \cite{RitortBJ05}. 
Throughout the paper we obtain the loading rate using $r_f=k_sv$. 

From the force distributions, computed at four different loading rates 
(Fig.\ref{P5GA_rf}-D),
we observe that the most probable rupture force ($f^*$) does not increase  logarithmically over a wide range of loading rates (Fig.\ref{P5GA_rf}-E).  Only if the range of $r_f$ is restricted $f^*$ changes linearly with $r_f$  \cite{Evans1}. 
The location of the transition state ($\Delta x_F^{TS}$) is usually
calculated using $f^* \approx (\frac{k_BT}{\Delta x_F^{TS}}) \log{r_f}$ 
\cite{Evans1} which may be reasonable as long as $r_f$ range is small. However, 
the $[f^*,\log{r_f}]$ plot is highly nonlinear (Fig. \ref{P5GA_rf}-E). 
If we use linear regression to analyze the $[f^*,\log{r_f}]$ plot then $\Delta 
x^{TS}_F\sim 0.8$ $nm$ for the distance
between the {\bf NBA} and the transition state. 
The small value of $\Delta x_F^{TS}$ is a consequence of the large variation of 
$\Delta x_F^{TS}$ as $r_f$ is changed \cite{HyeonPNAS05,RitortPRL06}.  
If the loading rate is varied over a broad range, the $r_f$-dependence of 
$\Delta x_{F}^{TS}$ is manifested as a pronounced convex curvature 
\cite{HyeonBJ06} 
in the $[f^*,\log{r_f}]$ plot (Fig. \ref{P5GA_rf}-E). 
Based on the equilibrium free energy profile $F(R)$ as a function of $R$ 
\cite{HyeonPNAS05} we expect that 
for P5GA $\Delta x_F^{TS}\approx 3$ $nm$ if $r_f$ is small. 
Indeed, from the constant force simulation results in Fig.\ref{P5GA_cf}-B we 
find $\Delta x_F^{TS}\approx \frac{R_U-R_F}{2}$. 
Thus, the slope of $[f^*,\log{r_f}]$ should decrease ($\Delta x_F^{TS}$ 
increases) 
as $r_f$ decreases. 
To illustrate the dramatic movement in the transition state we have calculate 
$\Delta x_F^{TS}$ using 
$f^*$ values for two consecutive values of $r_f$. 
For example, using $f^*\approx 12.5$ $pN$ at $r_f=450$ $pN/s$ and $f^*\approx 
14.9$ $pN$ at 
$r_f=4.5\times 10^3$ $pN/s$ (a value that can be realized in AFM experiments) 
we obtain $\Delta x_F^{TS}\approx 4.0$ $nm$. 
From the values of $f^*$ at five values of $r_f$ we find that $\Delta x_F^{TS}$ 
can move dramatically (Fig.\ref{P5GA_rf}-E). 
In particular, we find that $\Delta x_F^{TS}$ changes by nearly a factor of ten 
as the loading rate is decreased to 
values that are accessible in LOT experiments (see inset in 
Fig.\ref{P5GA_rf}-E).  Because of the nearly logarithmic variation of $\Delta x_F^{TS}$ on $r_f$ over a narrow range of $r_f$ we do not expect $\Delta x_F^{TS}$ to change appreciably if $r_f$ is lowered from $45 pN/s$ to $\approx 5pN/s$. 
At high $r_f$ or $f_c$ the unfolded state is greatly stabilized compared to the 
folded state. 
From Hammond's postulate \cite{HammondJACS53}, generalized to mechanical 
unfolding \cite{HyeonBJ06}, 
it follows that as $f_c$ increases $\Delta x_F^{TS}$ should move closer to the 
native state. 
The simulations are, therefore, in accord with the Hammond's postulate. 
\\

{\bf Force-induced transition in hairpins with bulges, internal loops - TAR RNA 
(PDB id : 1uud)}
The presence of bulges or internal loops contributes to the bending of the stiff 
helical stack. 
The enhanced flexibility enables formation of intramolecular tertiary contacts 
or facilitates 
RNA-protein interactions.  
Base-sugar, base-phosphate as well as base-base contacts are found in these 
structural elements in the native structure. 
For example, the base group of U23 in HIV-1 TAR RNA (PDB code \emph{1uud} 
\cite{DavisJMB04}, Fig.\ref{1uud_analysis}-A), 
protrudes away from the hairpin stack and makes tertiary contacts with $i=38-41$ 
(see three-dimensional structure in Fig.\ref{1uud_analysis}-B). 
The inherent flexibility in the bulge region also facilitates interaction with 
ligands \cite{WesthofARRBS97,WilliamsonNAR98}. 

To probe the effect of the U23C24U25 bulge (Fig.\ref{1uud_analysis}-A) on 
mechanical unfolding we performed force-ramp and force-clamp 
simulations. The time dependence of $R$ at $f_c=14$ $pN$, ({\bf NBA} is the 
preferred state) shows multiple transitions. 
Unlike in P5GA some of the transitions to {\bf UBA} involves a small pause in 
intermediate values 
of $R$ $(5-8)$ $nm$ which is suggestive of a short lived intermediate. 
Force-clamp simulations also show that $R$  fluctuates between 2 $nm$ ({\bf 
NBA}) 
and 10 $nm$ ({\bf UBA}). 
In addition, there is a signature for an intermediate with $R$ between $(5-8)$ 
$nm$. 
The free energy profile $\Delta F(R)$ shows that there is a high free energy 
intermediate centered at 
$R\approx 6$ $nm$ that is metastable with respect to the {\bf NBA} and the {\bf 
UBA}. 
At $f_c=14$ $pN$ the intermediate is less stable than the {\bf UBA} and the {\bf 
NBA}. 
It is likely that the instability of the intermediate state might make it 
difficult for experimental 
detection.
As force is increased to $f_c$ ($=14.7$ $pN$), which is very close to the 
critical value at which the 
stabilities of the folded and stretched states are nearly equal, the residence 
times in the {\bf NBA} and 
{\bf UBA} are similar (see middle panel in Fig. \ref{1uud_analysis}-D). 
The $P(R)$ distribution and $\Delta F(R)$ as a function of $R$ show that 
$f_c=14.7$ $pN$ 
is close to the critical value. 
Interestingly, at $f_c=14.7$ $pN$ the shallow high free energy intermediate is 
less pronounced than 
at $f_c=14.0$ $pN$ (Fig. \ref{1uud_analysis}-D). 
When force is further increased to $f_c=15.4$ $pN$ the intermediate becomes 
essentially a part of the {\bf UBA}. 
When the hairpins are unzipped the presence of bulges 
and internal loops contributes to the formation of the 
intermediate state when tertiary contacts between these bulges and 
the rest of the structure are disrupted. 
In the predicted intermediate the first six base pairs 
(Fig.\ref{1uud_analysis}-A) and additional contacts associated with these 
nucleotides are ruptured (see the representative structures in Fig. 
\ref{1uud_analysis}-D).

The contact map for TAR RNA has two ``clusters'' (Fig. \ref{1uud_analysis}-C). 
The one on the upper right corner that 
is associated with the lower hairpin stack and the other (nucleotides $25-31$) 
represents the structure 
from the bulge to the apical end. 
As the lower stack unfolds the force propagation goes along the diagonal for the 
upper cluster to 
the lower cluster (Fig.\ref{1uud_analysis}-C). 
We can qualitatively predict the kinetic barriers that oppose forced-unfolding 
at high $r_f$ 
using the contact map computed from the native structure (Fig. 
\ref{1uud_analysis}-C). 
If hairpin unfolding occurs in a \emph{sequential unzipping manner} then we 
expect (upon applying 
force to the 3' end) the structure to disrupt along the direction 
specified by the series of red arrows in Fig. \ref{1uud_analysis}-C. 
We predict that upon disruption of the first base pair (G17C45) the other 
contacts involving G17 and C45 
((17,39), (17,40), (17,41), (17,42), (17,43), (17,44), (17,45), (18,45), 
(19,45), and (20,45)) should 
spontaneously break (upper cluster). 
Similarly upon rupture of G18-C44 base pair the contacts associated with these 
nucleotides are disrupted (lower cluster). 
Therefore, the contact map suggests that rupture should occur as force 
propagates along the diagonal direction (red arrows in 
Fig.\ref{1uud_analysis}-C). 
When the contacts associated with the lower clusters unravel a high energy 
intermediate is 
populated (Fig. \ref{1uud_analysis}-C). 
Only detailed simulations can reveal the stability and lifetime of the 
intermediate. 

The free energy profile as a function of $R$ can also be computed from the contact 
map using
$\Delta F(R)=-N_c\epsilon_h\times\langle N_c(T=300K)\rangle/N_G-N_d\times (k_BTl_{ss}/d)\log{\left[\frac{\sinh(fd/k_BT)}{(fd/k_BT)}\right]}$  $\langle N_c(T=300K)\rangle$ is the average number of contacts at temperature $T=300K$ at zero force, 
and $N_G$ is the corresponding quality in the PDB structure.
The first term is the energetic contribution arising from $N_c$ surviving contacts and the second term accounts  for the entropy arising from the segment in which $N_d$ contacts are disrupted. 
The energetic contribution using $N_d$ and the entropy is given by the product of $N_d$ and the  entropy associated with the  freely jointed chain like model. We used Kuhn length $d\sim$ 2.0 $nm$, and the effective nucleotide length of single strand RNA $l_{ss}\sim 0.59$ $nm$. The chain extension is given by 
$R=R_0+N_dl_{ss}$ where $R_0$ is the end-to-end distance in the folded state. 
In the presence of constant force the free energy profile tilts to {\bf UBA}. 
The free energy profile $\Delta F(R)$ at $f_c \approx 17 pN$ $pN$ is suggestive of 
an intermediate $R\approx 6$ $nm$ whereas at higher force the signature of the 
intermediate disappears (see also Fig. \ref{1uud_analysis}-D). 
Since we approximated the conformational entropy using freely jointed chain 
model in this exercise,  
the estimate for the equilibrium critical force does not coincide with the  SOP simulations which shows that the 
freely jointed chain model does not estimate the entropy of the finite-sized RNA structures. 
Nevertheless, for simple hairpins the number of kinetic barriers or kinetic 
intermediates can be predicted using this simple analysis 
based on the only knowledge of native contact topology. 
More recently, 
Cocco \emph{et. al.} \cite{MarkoEPJE03} have proposed a similar scheme using the Monte Carlo simulations 
on the sequence dependent free energy profile computed with the Turner's 
thermodynamic rule \cite{MathewsJMB99}. 
A similar analysis was previously used to estimate equilibrium unbinding force 
for proteins \cite{Klimov}. 

The FECs obtained using force-ramp simulations at $r_f=4.5\times 10^3$ $pN/s$ 
(Fig. \ref{1uud_analysis}-E) 
also show that one intermediate is present. 
At this value of $r_f$ the presence of an intermediate occurs as a rip in the 
FECs (indicated using an arrow) at $f\approx 23$ $pN$. 
However, when $r_f$ is lowered by a factor of 10 (see right panel in Fig. 
\ref{1uud_analysis}-E) 
there is no signature of a rip at $f\approx 15-17$ $pN$ 
that corresponds to a pause in a high free energy metastable intermediate. 
As time increases the global unfolding are preceded by fluctuations between metastable intermediate to 
the folded state, and 
TAR RNA unfolds in an all-or-none manner at a force of about 20 $pN$. 
The picture that emerges from force-ramp simulations, namely, 
the presence of an intermediate at high $r_f$ and its absence at 
low $r_f$, is completely consistent with constant force simulations. 
The $r_f$-dependent rupture of RNA structures is a general property of 
self-organized molecules which we explore fully using physical arguments and 
simulations of ribozymes (see below). 
\\

\textbf{HCV IRES domain II:} 
We consider mechanical unfolding of the 55-nt domain IIa of the Hepatitis C 
Viral (HCV) genome whose NMR structure \cite{PuglisiNSB03} 
is known (PDB code: 1p5m).
The secondary structure map of the hairpin  contains bulges and is capped by the 
UUCG tertraloop at the apical end (Fig \ref{1p5m_analysis}A). 
The domain IIa oligonucleotide adopts a distorted L-shaped structure 
\cite{PuglisiNSB03} (Fig. \ref{1p5m_analysis}-B) with a relatively 
flexible hinge bulge (A53-A57) that is stabilized by Mg$^{2+}$.
Just as for the TAR RNA the number 
of plausible kinetic intermediates (in the appropriate force regime) in the 
\textbf{NBA} $\rightarrow$ \textbf{UBA} transition and the range of force and 
$R$ values over 
which they occur can be anticipated using the contact map, which reflects the 
nature of the native fold. 
The contact map (Fig. \ref{1p5m_analysis}-C) shows that it can be partitioned 
into three distinct clusters that are spatially adjacent.  
Upon application of force to the 3' end the rupture of contacts associated with 
nucleotides G1 and C55 occurs and force propagates diagonally 
(see upper right corner of Fig. 3C). There is a change in the structure of 
contact map with the breaking of contacts involving the base pair (A8-U49) 
which signals the formation of the first intermediate. 
As a result force propagates along the diagonal associated with second cluster. 
Upon disruption of A8U49 base pair force propagates along the diagonal of the 
sub-contact map 
(blue line in Fig. \ref{1p5m_analysis}-C). 
The second intermediate is populated when all the base contacts involving this 
substructure unravel.
Similarly, for 1p5m we expect population of the third intermediate 
that opposes forced unfolding associated with 
substructure involving contacts associated with the nucleotides that are in the 
vicinity of the green line in Fig. \ref{1p5m_analysis}-C.

Explicit force-clamp simulations confirm that there indeed are three 
intermediates associated with mechanical unfolding of \emph{1p5m}.  
We generated ten unfolding trajectories for a cumulative $800 ms$ at $f_c = 21 
pN$. 
As indicated approximately by the energy profile (Fig. \ref{1p5m_analysis}-C) we 
find 
multiple transitions between the structures that are revealed as plateaus in 
$R(t)$ (Fig. \ref{1p5m_analysis}-D).  
Not all the possible transitions are explicitly observed in all the 
trajectories. This observation may 
be a reflection of the intrinsic heterogeneity or stochastic nature of 
fluctuations. 
By averaging the residence time in each basin of attraction over time and the 
initial conditions we obtain $P(R)$ and the associated free energy profile (Fig. 
\ref{1p5m_analysis}-D).  
The \textbf{NBA} $\rightarrow$ \textbf{UBA} occurs through a sequence of three 
intermediates. 
The barrier separating the \textbf{UBA} and the first intermediate ($R \approx 5 
nm$) is larger than the subsequent ones. 
All the intermediates, whose populations are $f_c$ dependent (see below), are 
metastable with respect to 
{\bf NBA} and {\bf UBA}.  
The predicted intermediates can be detected experimentally provided they are long-lived. If the
lifetime of the metastable intermediate is too short then a high time resolution would be required.
Identical unfolding pathway is also found in force-ramp simulations (Fig. 
\ref{1p5m_analysis}-E) in which we find, in many trajectories, 
three rips that represent the kinetic intermediates. 
The forces at which these intermediates are populated are higher because the 
loading rate is large. 
The structures that are populated along the unfolding pathway (Fig. 
\ref{1p5m_analysis}-D) are in accord with the predictions 
based on the secondary structure and contact maps.  

Just as found explicitly for TAR RNA the intermediates may not be populated at 
lower $f_c$ or $r_f$ values. 
From the energy profile in Fig.\ref{1p5m_analysis}-C we find (approximately) 
that 
the energy associated with a putative intermediate at $R\approx 10$ $nm$ (see 
Fig. \ref{1p5m_analysis}-D) is 
$\Delta F(R\approx 10nm)\approx 50 k_BT\approx 200pN\cdot nm$. 
Thus, the smallest value of $f$ required to transiently populate the $R\approx 
10$ $nm$ intermediate is $f_I\approx \Delta F(R\approx 10nm)/10nm\approx 20pN$. 
Only if $f$ exceeds 20 $pN$ can these intermediates be significantly populated. 
At these forces the predicted intermediates are metastable 
with respect to both {\bf UBA} and {\bf NBA}.  
Their experimental detection would depend on their life times.
Unless relatively high time resolution is used in experiments, for practical purposes, mechanical unfolding of domain IIa of the HCV IR2S RNA might follow
 two-state behavior.
The simple consideration and the simulations might explain the reason for apparent absence 
of intermediates in the recent constant force LOT experiments on TAR RNA 
performed at 
$f<15$ $pN$.  
\\

{\bf Three-way junction prohead RNA (pRNA) : }
The next level of complexity in the secondary structural RNA motif 
is the one that contains an internal multiloop. 
We choose the prohead RNA (PDB code $1foq$ \cite{RossmanNature00}) which is part 
of the $\phi$29 DNA packaging motor. 
In the context of the motor pRNA assembles as a pentamer with each monomer 
consisting of 
two stem-loops that are separated by an internal multiloop 
(Fig.\ref{1FOQ_map}-A). 
Even at this level of complexity it becomes difficult to predict the kinetic 
barriers associated with force-induced
unfolding using the contact map alone. 
For the 109 nucleotide RNA, we generated four mechanical force-ramp unfolding 
trajectories (Fig.\ref{1FOQ_map}-B). 
The mechanical unfolding of the structure 
exhibits multiple rips signaling the presence of kinetic barriers separating the 
{\bf NBA} and {\bf UBA} (Fig.\ref{1FOQ_map}-B). 
The four different FECs, colored in black, red, green, and blue, are distinct. 
Variations in the rip dynamics from molecule to molecule may reflect the 
heterogeneous nature of the unfolding pathways. 
Upon force-ramp unfolding of the pRNA begins at $R\sim 2$ $nm$ in all the 
trajectories (Fig.\ref{1FOQ_map}-B). 
The FEC in red has three rips at $R\approx$15, 28, and 38 $nm$ while 
the FEC in green has only one rip at $R\approx$38 $nm$.
For pRNA, which is structurally more complex than the simpler three way junction 
(P5abc$\Delta$A for example \cite{Bustamante2}) the details of 
the unraveling mechanism is difficult to extract using the FECs alone. 
To unambiguously extract the unfolding pathways we have calculated the time 
dependence of nucleotide-dependent rupture (Fig. \ref{1FOQ_map}-C) of individual 
contacts (Methods). 

Because of the additional stem-loop that branches out of the internal multiloop 
(Fig.\ref{1FOQ_map}-A), 
there are potentially two routes by which the unfolding of pRNA can proceed. 
One is  $P1\Rightarrow P3\Rightarrow P2$, and the other is  $P1\Rightarrow 
P2\Rightarrow P3$. 
We can rule out the unfolding pathways such as $P2\Rightarrow P1\Rightarrow P3$ 
or 
$P3\Rightarrow P2\Rightarrow P1$ because of the directional nature of the 
applied mechanical force. 
It should be stressed that such pathways can emerge if force is applied at 
points other than the ends of the chain. 
Using the time evolution of the native contact of the $i^{th}$ nucleotide, 
$Q_i(t)$, we 
unambiguously pin down the time series of the rupture of individual contacts 
upon force-ramp (Fig.\ref{1FOQ_map}-C). 
The rupture pattern of the individual contacts in all the trajectories is very 
similar (Fig.\ref{1FOQ_map}-C) despite seeming differences in 
the FECs. 
From Fig.\ref{1FOQ_map}-C it is clear that the nucleotides in P1 unravel 
($Q_i(t))$ associated with these nucleotides are zero for $t>5ms$)
early (around $t\sim 5ms$) which signals the first event 
in pRNA unfolding. 
Subsequently, $Q_i(t)$ for $i\sim(70-85)$ are disrupted. 
This shows that the smaller stem-loop in P3 opens, forming a cusp at $R\sim 15$ 
$nm$ in the FEC. 
The bigger stem-loop with the large bulge in P2 follows the opening of P3 which 
is manifested as a 
rip near $R\sim 35$ $nm$ in the FEC. 
Thus, the unfolding pathway for pRNA is $P1\Rightarrow P3\Rightarrow P2$.
Depending on the trajectory, not all the three rips are detected in the FECs 
although the unfolding pathway defined by the subdomains 
is not altered. 
The sequence of structural changes that accompany the unfolding of pRNA for the 
trajectory in red in  Fig.\ref{1FOQ_map}-B is shown in Fig.\ref{1FOQ_map}-D. 
Force-ramp simulations of pRNA show that even when the native structure is not 
too complex it is difficult to predict the nature of kinetic barriers without 
detailed dynamical simulations or experiments. 
It would be interesting to test these predictions using LOT experiments. \\

\section{CONCLUSIONS}

We have used the self-organized polymer representation of a variety of RNA structures to predict their mechanical unfolding trajectories.  
Constant force and force-ramp simulations show that dramatic changes in the force profiles take place as the loading rates and the values of the force are varied.  
If the force is varied over a wide range then regions of the energy landscape that cannot be accessed in conventional experiments can be probed.  
However, in order to realize the full utility of the single molecule force spectroscopy it becomes necessary to use force in distinct modes 
(constant force, force-ramp, and other combinations) along with reliable computations that can mimic the experimental conditions as closely as possible. 
The simulations of RNA, with diverse native structures,  using the SOP model illustrate the structural details in the unfolding pathways that are experimentally accessible.  
It is remarkable that the simple native state-based SOP model can quantitatively predict the FECs for a number of RNA molecules with varying degree of structural
complexity.  
We conclude the paper with the following additional remarks.

\textit{Transition state movements show changes from plastic to brittle behavior:}  
The small size and simple architecture of RNA hairpins has allowed us to explore their 
response to force over a range of $r_f$ that spans four orders of magnitude. The lowest $r_f$ value is close to those used in LOT experiments.  
A key prediction of our simulations is that the location of the transition state for P5GA moves dramatically from about 6 $nm$ at low $r_f$ to about 0.5 $nm$ at high $r_f$ (see inset to Fig. 5E).  
The large value of $\Delta x_F^{TS}$ at low $r_f$ suggests that P5GA is plastic while the small $\Delta x_F^{TS}$ at high $r_f$ is suggestive of brittle behavior.  
The mechanical properties of RNA structures can be drastically altered by varying the loading rate
which is reminiscent of the changes in the visco-elastic behavior of polymeric materials that changes with frequency.  
The transformation from plastic to brittle behavior can be captured using the fragility index \cite{RitortPRL06} used to describe mechanical unfolding of hairpins.    
Although we have discussed the $r_f$-dependent movement of the transition states using P5GA as an example 
we predict that this result is general and should be observed in other RNA structures as well.

The predicted movement in the transition state as a function of $r_f$ might help resolve the apparent differences in the estimated values of $\Delta x_F^{TS}$ 
in proteins using LOT  and AFM experiments.  The typical value of $\Delta x_F^{TS}$ for a number of proteins using AFM is on the order of 0.5 $nm$ \cite{GaubSCI97}
whereas $\Delta x_f^{TS}$ for RNase H \cite{MarquseeScience05} (the only protein that has been experimentally studied using the LOT setup) is about 6$nm$.  
We believe that such a large difference is not merely due to changes in native topology and stability, which undoubtedly are important. 
Rather, it is due to the variations in $r_f$.  The value of $r_f$ in LOT is typically less than 10$pN/s$ whereas $r_f$ in AFM varies from (100 - 1000)$pN/s$.  
Our findings here suggest that the different loading rates used in the two setups might explain the large differences in the values of $\Delta x_F^{TS}$. 
Thus, it is important to perform experiments on a given protein using both LOT and AFM setup to sort out 
the loading rate dependence of the location of the transition state.

\textit{Forced-unfolding pathways for simple RNA are encoded in the contact map:}  From the contact map it is possible to
visualize the directions along which the applied tension propagates.  As illustrated using simple RNA structures the unfolding
pathway depends on the direction of tension propagation and local structural stability.  Using the contact map alone
one can anticipate the most probable unfolding pathways.  Here, we have shown that for P5GA, TAR RNA, and HCV IRES domain II
it is possible to get a qualitative picture of forced-unfolding using the reduced representation of RNA structures in the form of  contact maps.  However,
as the structural complexity increases and a number of alternate unfolding pathways become possible the simple native structure based method
alone is not always sufficient in predicting how a particular structure unravels.  Such is the case in the unfolding of
the three-way junction prohead RNA studied here at constant $r_f$.    

\textit{Limitations of the SOP model:}  The SOP model is remarkably successful 
in reproducing the unfolding pathways of complex ribozymes in a realistic fashion.
Surprisingly, for both proteins \cite{HyeonSTRUCTURE06} and RNA, the SOP model is quite successful in predicting the nature of unfolding pathways.   
The unfolding dynamics of proteins as well as RNA, with size exceeding 250 residues or nucleotides  
can be conveniently simulated on a PC in a few days with the simulation 
condition used in this paper ($r_f\approx 10^2-10^5$ $pN/s$).
Quenching the force to zero drives the stretched state of molecule close to the native state \cite{HyeonSTRUCTURE06}, which allows us to map the folding pathways starting from different initial conditions.  
The SOP model consisting of the polymeric nature and the minimal characteristics 
of RNA architecture is reasonable in visualizing the forced-unfolding and force-quench refolding dynamics of RNA molecules. 
We believe that, when accompanied with the experimental analysis, 
the SOP model can serve as a useful tool that provides insights into the 
folding/unfolding process of large macromolecules. 
However, like all models there are certain limitations of the SOP model which prevent us from making
quantitative predictions of the measurable force-extension curves especially for large RNA.  
The sequence and/or the counterion effect are not explicitly taken into 
account in the SOP model. 
The neglect of explicit  counterions in the simulations, fails to capture their  specific coordination with RNA,  which in turn leads to an underestimate of the local stability of the
folded structure.  Hence in applying the SOP model to investigate structures in which counterion-mediated tertiary interactions determine
local structures  $\epsilon_h$ should be varied. Despite the obvious limitations it is clear that the SOP model is powerful enough to provide
insights into the structures and pathways that are explored upon application of force. The model can not only be used as a predictive tool (as 
shown here with explicit applications on systems for which experiments are not currently available) but also can be used to
interpret experimental results.

{\bf Acknowledgements:} 
We are grateful to Prof. Ruxandra I. Dima and Dr. David Pincus for a number of insightful comments. 
This work was supported in part by a grant from the National Science Foundation 
through grant number NSF CHE-05-14056.

\newpage 

\section{Figure Captions}

\begin{figure}[ht]
\includegraphics[width=5.00in]{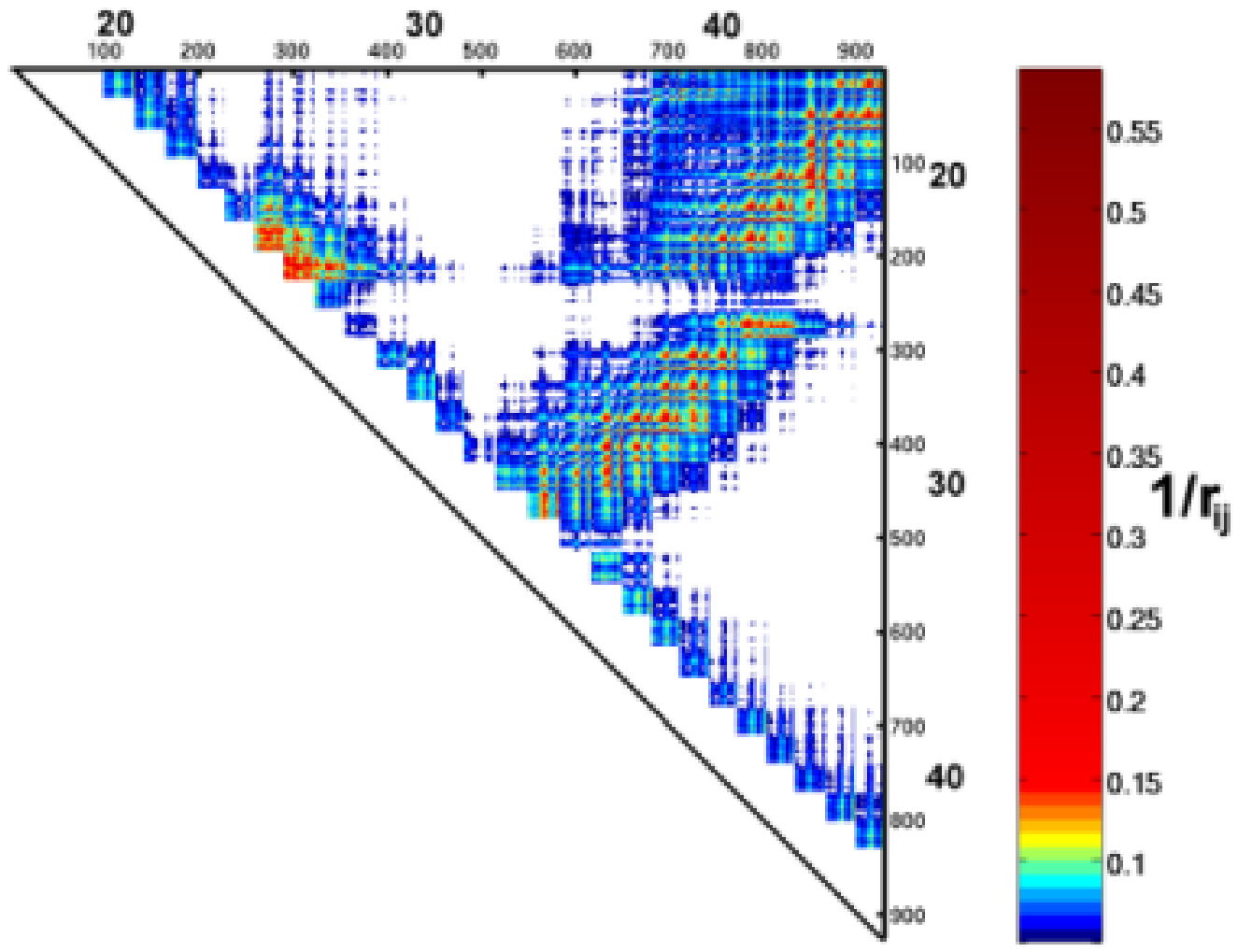} 
\caption{\label{distancemap_1uud}}
\end{figure}

{{\bf Figure} \ref{distancemap_1uud} :}
The distance map for TAR RNA (PDB code 1uud). 
The distance ($r_{ij}$ [\AA]) between all the heavy atoms (C, N, O, P) are 
computed for the nucleotides ($|i-j|>2$ with 
$r_{ij}<2$ $nm$). 
The scale on the right gives the inverse of distances ($1/r_{ij}$) in color. 
Small rectangular lattices corresponds to the nucleotide unit. 
The index for the heavy atoms is labeled in small font on $i$ and $j$ axis, and 
the index of the nucleotides ($i=17-45$) 
is labeled using a large font.
Upon coarse graining using the SOP model we obtain the contact map in Fig. 
\ref{1uud_analysis}-C.

\begin{figure}[ht]
\includegraphics[width=6.00in]{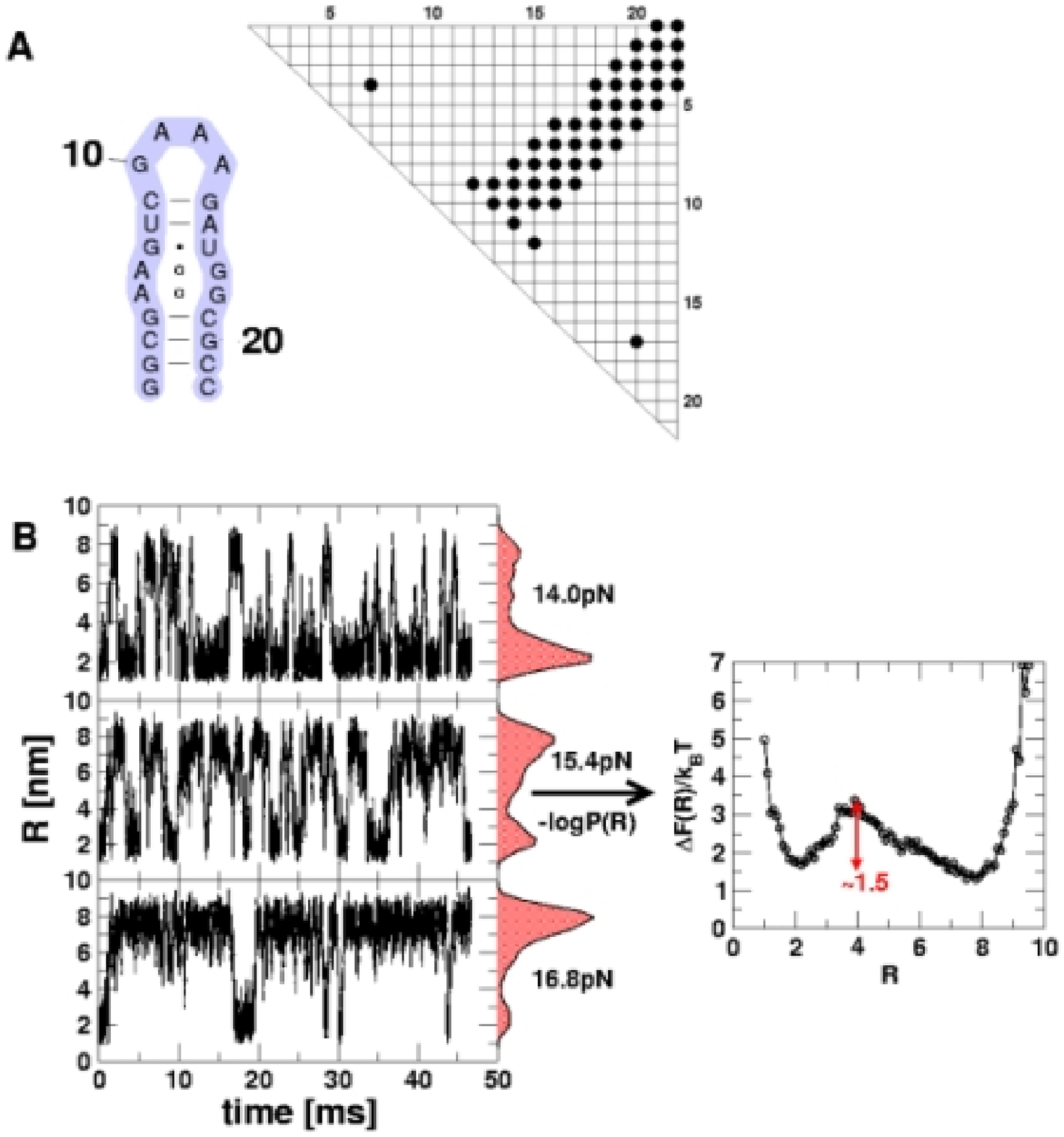} 
\caption{\label{P5GA_cf}}
\end{figure}

{{\bf Figure} \ref{P5GA_cf} :}
{\bf A.} The secondary structure of P5GA hairpin and its contact map. 
{\bf B.}
The time-dependent fluctuations of P5GA hairpin between the folded ($R\approx 
1.5 nm$) and unfolded ($R\approx 8nm$) states. 
The end-to-end distance changes spontaneously between two values. 
The force-clamped dynamics of the P5GA hairpin is probed for $\sim 45$ $ms$.
The histograms $P(R)$ at $f_c=14.0$, $15.4$, $16.8$ $pN$ are shown in red. 
The free energy profile $\Delta F(R)$ as a function of $R$ for $f_c=15.4$ $pN$ 
on the right 
shows two-state behavior. 
 
\begin{figure}[ht]
\includegraphics[width=6.00in]{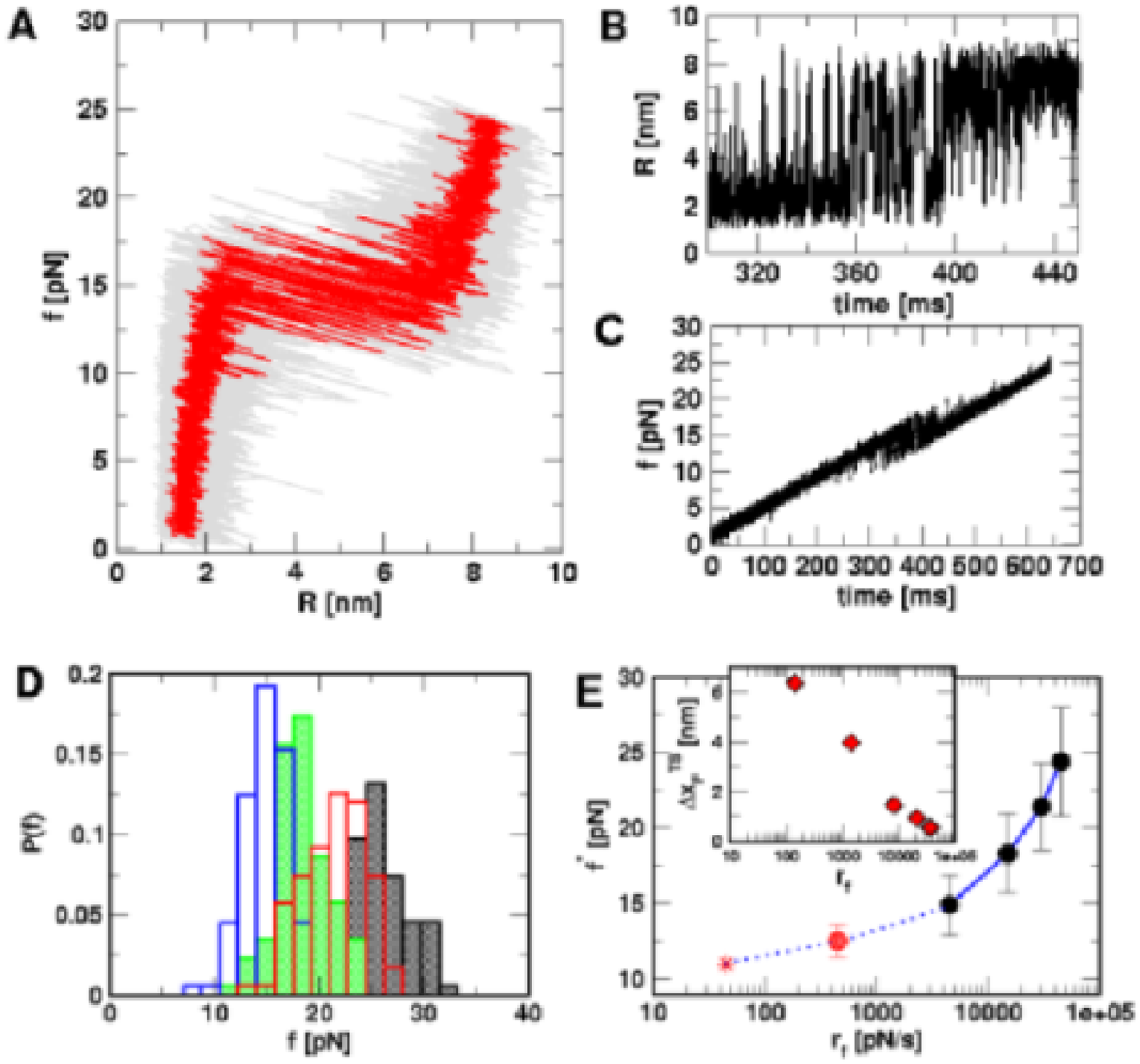} 
\caption{\label{P5GA_rf}}
\end{figure}

\begin{figure}[ht]
\includegraphics[width=7.00in]{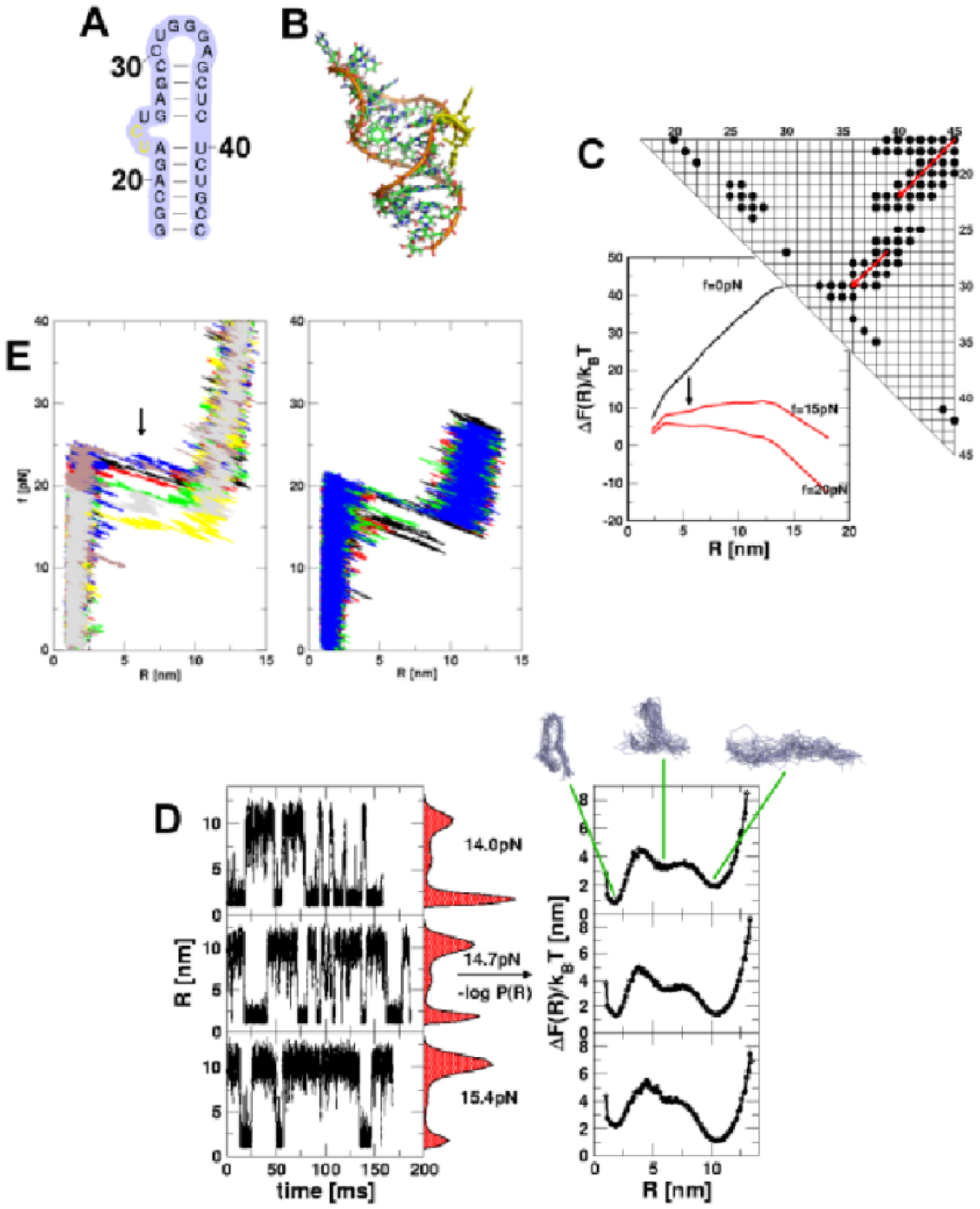} 
\caption{\label{1uud_analysis}}
\end{figure}

{{\bf Figure} \ref{P5GA_rf} :}
Force-ramp unfolding of P5GA hairpin. 
{\bf A.}
An example of FEC at the loading rate $r_f=45$ $pN/s$ ($k_s=0.07$ $pN/nm$, 
$v=0.64$ $\mu m/s$).
The data points are recorded every 50 $\mu s$ (grey color), but for better 
illustration the running average is displayed every 
500 $\mu s$ (red color). 
{\bf B.} The end-to-end distance ($R$) as a function of time. 
{\bf C.} Time dependence of $f$. The loading rate $r_f=df/dt$ is nearly a 
constant. 
{\bf D.} Distribution of unbinding forces from 100 trajectories at four loading 
rates 
$r_f=r_f^o$, $\frac{10}{3} r_f^o$, $\frac{20}{3} r_f^o$, and $10 r_f^o$ 
($r_f^o=4.5\times 10^3$ $pN/s$ ($k_s=0.7$ $pN/nm$, $v=6.4$ $\mu m/s$)). 
The red star is the rupture force value ($f^*=12.5$ $pN$) at $r_f=45$ $pN/s$.
{\bf E.} Plot $f^*$, the most probable unfolding force, as a function of $r_f$. 
The position of transition state $\Delta x_F^{TS}$ is computed using 
$\Delta x_F^{TS}=k_BT\frac{\Delta\log{r_f}}{\Delta f^*}$.
The inset shows the variation of $\Delta x_F^{TS}$ as a function of $r_f$.

{{\bf Figure} \ref{1uud_analysis} :}
Analysis of force-induced transitions in TAR RNA hairpin. 
{\bf A.} Secondary structure. 
{\bf B.} In the three dimensional structure 
the nucleotides ($i=23,24$) in the bulge are in yellow color. 
{\bf C.} The contact map and the energy profile at three values of $f$. 
The plot suggests the presence of one intermediate at $R \sim 6$ $nm$ when 
$f\neq 0$. 
{\bf D.} The results of force-clamp simulations at $f_c=14$ $pN$, 14.7 $pN$ and 
15.4 $pN$. 
The dynamics of $R$ is
probed for 150 $ms$ and the histograms $P(R)$ is given in the shaded color. 
The corresponding free energy profiles computed using $\Delta 
F(R)/k_BT=-\log{P(R)}$ are shown on the right.
The structures that correspond to the three basins of attraction are displayed 
at the top.
{\bf E.} FECs at the loading  rates, $r_f^o=4.5\times 10^3$ $pN/s$ ($k_s=0.7$ 
$pN/nm$, $v=6.4$ $\mu m/s$) (left) and 
$r_f^o=4.5\times 10^2$ $pN/s$ ($k_s=0.7$ $pN/nm$, $v=0.64$ $\mu m/s$) (right) 
for a number of molecules. 
The arrow in the left panel is the signature of a kinetic intermediate which is 
absent when $r_f$ is reduced. 

\begin{figure}[ht]
\includegraphics[width=7.50in]{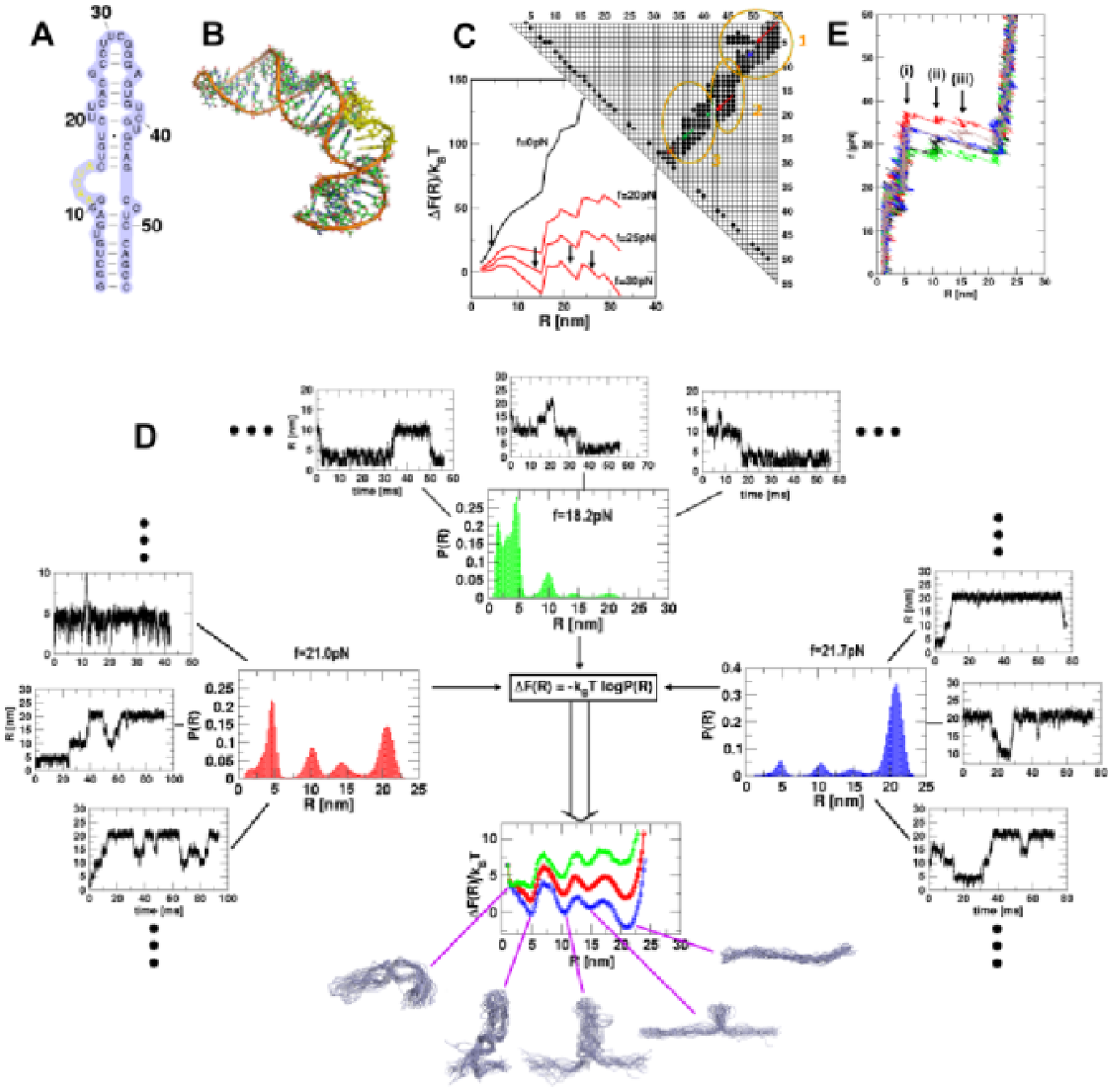} 
\caption{\label{1p5m_analysis}}
\end{figure}

{{\bf Figure} \ref{1p5m_analysis} :}
Force-induced transitions in domain IIa of HCV IRES RNA. 
{\bf A.} Secondary structure. {\bf B.} The nucleotides ($i=12-16$) in bulge in 
the 
three dimensional structure are in yellow. 
{\bf C.} The contact map and the associated energy profile as a function of $R$ 
at three values of $f_c$.  The three clustered regions are encircled for clarity.
The presence of three metastable intermediates at $R \sim$5, 15, 22, 28 $nm$ is 
indicated by arrows. 
{\bf D.} The results of force-clamp simulations at $f_c=$18.2, 21.0, and 21.7 $pN$. 
The multiple dynamical transitions in the hairpin is probed by following the 
time dependence of $R$. For each force the transitions, which are probed for 800 $ms$ duration 
in 10 trajectories, 
are shown for three molecules. 
The distributions $P(R)$ averaged over time and the initial conditions (the dots 
indicate seven other trajectories) are shown next to the displayed trajectories and the free energy 
profile $\Delta F(R)$ is shown below. The color in the $\Delta F(R)$ profiles are the same as in $P(R)$ at each force.
The structures of the intermediates, the native conformation, and the unfolded 
structures are explicitly shown. 
{\bf E.} FECs at the loading  rate, $r_f^o=4.5\times 10^3$ $pN/s$ ($k_s=0.7$ 
$pN/nm$, $v=6.4$ $\mu m/s$) for three trajectories.  
The arrow indicates the kinetic intermediates. 

\begin{figure}[ht]
\includegraphics[width=6.0in]{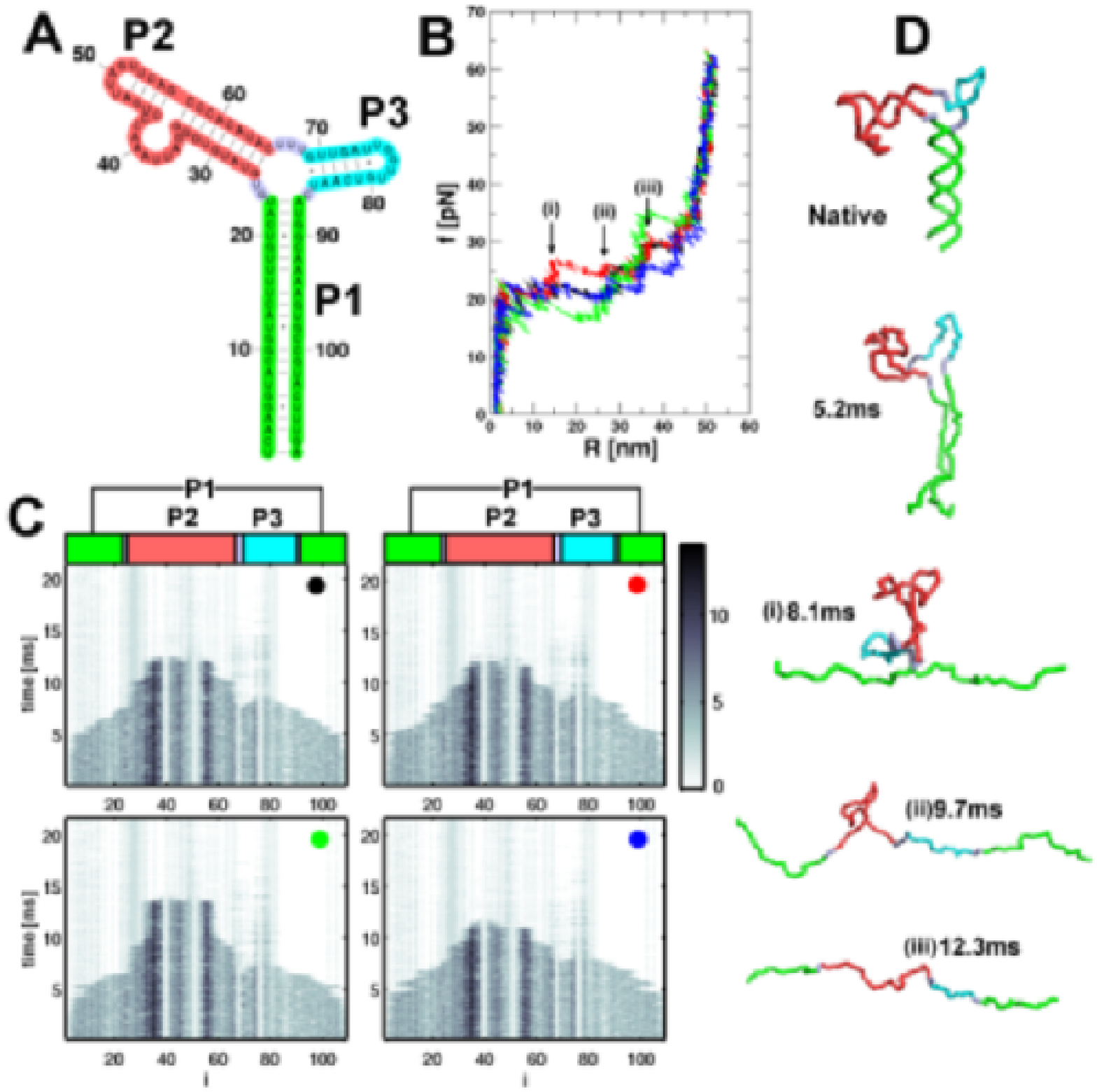} 
\caption{\label{1FOQ_map}}
\end{figure}

{{\bf Figure} \ref{1FOQ_map} :}
{\bf A.} The secondary structure of 109 nt pRNA.
{\bf B.} The four FECs of pRNA at $r_f^o=4.5\times 10^3$ $pN/s$ ($k_s=0.7$ 
$pN/nm$, $v=6.4$ $\mu m/s$).
{\bf C.} The rupture history of pRNA ($Q_i(t)$) corresponding to FECs in {\bf 
B}. 
Here $Q_i(t)$ is the number of contacts that the nucleotide $i$ has at time $t$. 
The scale on the right represents the number of contacts associated with the 
$i^{th}$ nucleotide. 
Darker shades have larger number of contacts than lighter shades. 
{\bf D.} Structures of pRNA at each stage of unfolding as time progresses is 
illustrated. 
Structures in (i)$-$(iii) are associated with kinetic barriers in the red 
trajectory in {\bf B}.

\newpage

\begin{thebibliography}{44}
\providecommand{\url}[1]{\texttt{#1}}
\providecommand{\urlprefix}{ }

\bibitem[Cech et~al.(1981)Cech, Zaug, and Grabowski]{CechCell81}
Cech, T.~R., A.~J. Zaug, and P.~J. Grabowski, 1981.
\newblock \emph{In vitro} splicing of the ribosomal-{RNA} precursor of
  \emph{Tetrahymena}-involvement of a quianosine nucleotide in the excision of
  the intervening sequence.
\newblock \emph{Cell} 27:487--496.

\bibitem[Guerrier-Takada and Altman(1984)]{AltmanSci84}
Guerrier-Takada, C., and S.~Altman, 1984.
\newblock Catalytic activity of an {RNA} molecule prepared by transcription
  \emph{in vitro}.
\newblock \emph{Science} 223:285--286.

\bibitem[Doudna and Cech(2002)]{DoudnaNature02}
Doudna, J., and T.~Cech, 2002.
\newblock The chemical repertoire of natural ribozymes.
\newblock \emph{Nature} 418:222--228.

\bibitem[Treiber and Williamson(2001)]{TreiberCOSB01}
Treiber, D.~K., and J.~R. Williamson, 2001.
\newblock Beyond kinetic traps in {RNA} folding.
\newblock \emph{Curr. Opin. Struct. Biol.} 11:309--314.

\bibitem[Sosnick and Pan(2003)]{SosnickCOSB03}
Sosnick, T., and T.~Pan, 2003.
\newblock {RNA} folding: models and perspectives.
\newblock \emph{Curr. Opin. Struct. Biol.} 13:309--316.

\bibitem[Pan et~al.(1997)Pan, Thirumalai, and Woodson]{PanJMB97}
Pan, J., D.~Thirumalai, and S.~A. Woodson, 1997.
\newblock Folding of {RNA} involves parallel pathways.
\newblock \emph{J. Mol. Biol.} 273:7--13.

\bibitem[Thirumalai and Hyeon(2005)]{HyeonBC05}
Thirumalai, D., and C.~Hyeon, 2005.
\newblock {RNA and Protein folding: Common Themes and Variations}.
\newblock \emph{Biochemistry} 44:4957--4970.

\bibitem[Bokinsky and Zhuang(2005)]{ZhuangACR05}
Bokinsky, G., and X.~Zhuang, 2005.
\newblock Single-molecule {RNA} folding.
\newblock \emph{Acc. Chem. Res.} 38:566--573.

\bibitem[Russell and Herschlag(1999)]{HerschlagJMB99}
Russell, R., and D.~Herschlag, 1999.
\newblock New pathways in folding of the \emph{Tetrahymena} group {I} {RNA}
  enzyme.
\newblock \emph{J. Mol. Biol.} 291:1155--1167.

\bibitem[Zhuang et~al.(2000)Zhuang, Bartley, Babcock, Russell, Ha, Hershlag,
  and Chu]{ZhuangSCI00}
Zhuang, X., L.~Bartley, A.~Babcock, R.~Russell, T.~Ha, D.~Hershlag, and S.~Chu,
  2000.
\newblock A single-molecule study of {RNA} catalysis and folding.
\newblock \emph{Science} 288:2048--2051.

\bibitem[Liphardt et~al.(2001)Liphardt, Onoa, Smith, {Tinoco, Jr.}, and
  Bustamante]{Bustamante2}
Liphardt, J., B.~Onoa, S.~B. Smith, I.~{Tinoco, Jr.}, and C.~Bustamante, 2001.
\newblock {Reversible unfolding of single {RNA} molecules by mechanical force}.
\newblock \emph{Science} 292:733--737.

\bibitem[Onoa et~al.(2003)Onoa, Dumont, Liphardt, Smith, {Tinoco, Jr.}, and
  Bustamante]{Bustamante4}
Onoa, B., S.~Dumont, J.~Liphardt, S.~B. Smith, I.~{Tinoco, Jr.}, and
  C.~Bustamante, 2003.
\newblock {Identifying Kinetic Barriers to Mechanical Unfolding of the \emph{T.
  thermophila} Ribozyme}.
\newblock \emph{Science} 299:1892--1895.

\bibitem[Bustamante et~al.(1994)Bustamante, Marko, Siggia, and
  Smith]{BustamanteSCI94}
Bustamante, C., J.~F. Marko, E.~D. Siggia, and S.~Smith, 1994.
\newblock Entropic Elasticity of $\lambda$-Phase {DNA}.
\newblock \emph{Science} 265:1599--1600.

\bibitem[Marko and Siggia(1996)]{MarkoMacro96}
Marko, J.~F., and E.~D. Siggia, 1996.
\newblock {Bending and Twisting Elasticity of DNA}.
\newblock \emph{Macromolecules} 27:981--988.

\bibitem[Jarzynski(1997)]{JarzynskiPRL97}
Jarzynski, C., 1997.
\newblock Nonequilibrium Equality for Free Energy Differences.
\newblock \emph{Phys. Rev. Lett.} 78:2690.

\bibitem[Liphardt et~al.(2002)Liphardt, Dumont, Smith, {Tinoco, Jr.}, and
  Bustamante]{Bustamante3}
Liphardt, J., S.~Dumont, S.~B. Smith, I.~{Tinoco, Jr.}, and C.~Bustamante,
  2002.
\newblock {Equilibrium information from nonequilibrium measurements in an
  experimental test of Jarzynski's equality}.
\newblock \emph{Science} 296:1832--1835.

\bibitem[Hyeon and Thirumalai(2005)]{HyeonPNAS05}
Hyeon, C., and D.~Thirumalai, 2005.
\newblock {Mechanical unfolding of RNA hairpins}.
\newblock \emph{Proc. Natl. Acad. Sci.} 102:6789--6794.

\bibitem[Li et~al.(2006)Li, Collin, Smith, Bustamante, and {Tinoco,
  Jr.}]{TinocoBJ06}
Li, P. T.~X., D.~Collin, S.~B. Smith, C.~Bustamante, and I.~{Tinoco, Jr.},
  2006.
\newblock Probing the Mechanical Folding Kinetics of TAR RNA by Hopping,
  Force-Jump, and Force-Ramp Methods.
\newblock \emph{Biophys. J.} 90:250--260.

\bibitem[Fernandez and Li(2004)]{FernandezSCI04}
Fernandez, J.~M., and H.~Li, 2004.
\newblock {Force-Clamp Spectroscopy Monitors the Folding Trajectory of a Single
  Protein}.
\newblock \emph{Science} 303:1674--1678.

\bibitem[Carrion-Vazquez et~al.(1999)Carrion-Vazquez, Oberhauser, Fowler,
  Marszalek, Broedel, Clarke, and Fernandez]{FernandezPNAS99}
Carrion-Vazquez, M., A.~F. Oberhauser, S.~B. Fowler, P.~E. Marszalek, S.~E.
  Broedel, J.~Clarke, and J.~M. Fernandez, 1999.
\newblock Mechanical and chemical unfolding of a single protein: {A}
  comparison.
\newblock \emph{Proc. Natl. Acad. Sci.} 96:3694--3699.

\bibitem[Gerland et~al.(2003)Gerland, Bundschuh, and Hwa]{HwaBP03}
Gerland, U., R.~Bundschuh, and T.~Hwa, 2003.
\newblock {Mechanically Probing the Folding Pathway of Single RNA Molecules}.
\newblock \emph{Biophys. J.} 84:2831--2840.

\bibitem[Gerland et~al.(2001)Gerland, Bundschuh, and Hwa]{HwaBP01}
Gerland, U., R.~Bundschuh, and T.~Hwa, 2001.
\newblock {Force-Induced Denaturation of RNA}.
\newblock \emph{Biophys. J.} 81:1324--1332.

\bibitem[Cocco et~al.(2003)Cocco, Marko, and Monasson]{MarkoEPJE03}
Cocco, S., J.~Marko, and R.~Monasson, 2003.
\newblock Slow nucleic acid unzipping kinetics from sequence-defined barriers.
\newblock \emph{Eur. Phys. J. E} 10:153--161.

\bibitem[Hyeon and Thirumalai(2006)]{HyeonBJ06}
Hyeon, C., and D.~Thirumalai, 2006.
\newblock Forced-unfolding and force-quench refolding of {RNA} hairpins.
\newblock \emph{Biophys. J.} 90:3410--3427.

\bibitem[Klimov and Thirumalai(2000)]{Klimov2}
Klimov, D.~K., and D.~Thirumalai, 2000.
\newblock Native topology determines force-induced unfolding pathways in
  globular proteins.
\newblock \emph{Proc. Natl. Acad. Sci. USA} 97:7254--7259.

\bibitem[Kremer and Grest(1990)]{KremerJCP90}
Kremer, K., and G.~S. Grest, 1990.
\newblock Dynamics of entangled linear polymer melts: A molecular-dynamics
  simulation.
\newblock \emph{J. Chem. Phys.} 92:5057--5086.

\bibitem[Hyeon et~al.(2006)Hyeon, Dima, and Thirumalai]{HyeonSTRUCTURE06}
Hyeon, C., R.~I. Dima, and D.~Thirumalai, 2006.
\newblock Pathways and kinetic barriers in mechanical unfolding and refolding
  of RNA and proteins.
\newblock \emph{Structure (in press)} .

\bibitem[Ermak and {McCammon}(1978)]{McCammonJCP78}
Ermak, D.~L., and J.~A. {McCammon}, 1978.
\newblock Brownian dynamics with hydrodynamic interactions.
\newblock \emph{J. Chem. Phys.} 69:1352--1369.

\bibitem[Veitshans et~al.(1996)Veitshans, Klimov, and
  Thirumalai]{VeitshansFoldDes96}
Veitshans, T., D.~Klimov, and D.~Thirumalai, 1996.
\newblock Protein folding kinetics: timescales, pathways and energy landscapes
  in terms of sequence-dependent properties.
\newblock \emph{Folding Des.} 2:1--22.

\bibitem[Rudisser and {Tinoco, Jr.}(2000)]{TinocoJMB2000}
Rudisser, S., and I.~{Tinoco, Jr.}, 2000.
\newblock {Solution Structure of Cobalt(III)Hexammine Complexed to the GAAA
  Tetraloop, and Metal-ion Binding to GA Mismatches}.
\newblock \emph{J. Mol. Biol.} 295:1211--1223.

\bibitem[Hofacker(2003)]{HofackerNAR03}
Hofacker, I.~V., 2003.
\newblock Vienna {RNA} secondary structure server.
\newblock \emph{Nucl. Acids. Res.} 31:3429--3431.

\bibitem[Manosas and Ritort(2005)]{RitortBJ05}
Manosas, M., and F.~Ritort, 2005.
\newblock {Thermodynamic and Kinetic Aspects of RNA pulling experiments}.
\newblock \emph{Biophys. J.} 88:3224--3242.

\bibitem[Evans and Ritchie(1997)]{Evans1}
Evans, E., and K.~Ritchie, 1997.
\newblock {Dynamic Strength of Molecular Adhesion Bonds}.
\newblock \emph{Biophys. J.} 72:1541--1555.

\bibitem[Manosas et~al.(2006)Manosas, Collin, and Ritort]{RitortPRL06}
Manosas, M., D.~Collin, and F.~Ritort, 2006.
\newblock Force-Dependent Fragility in {RNA} Hairpins.
\newblock \emph{Phys. Rev. Lett.} 96:218301.

\bibitem[Hammond(1953)]{HammondJACS53}
Hammond, G.~S., 1953.
\newblock A correlation of reaction rates.
\newblock \emph{J. Am. Chem. Soc.} 77:334--338.

\bibitem[Davis et~al.(2004)Davis, Afshar, Varani, Murchie, Karn, Lentzen,
  Drysdale, Bower, Potter, Starkey, Swarbrick, and Aboul-ela]{DavisJMB04}
Davis, B., M.~Afshar, G.~Varani, A.~I.~H. Murchie, J.~Karn, G.~Lentzen,
  M.~Drysdale, J.~Bower, A.~J. Potter, I.~D. Starkey, T.~Swarbrick, and
  F.~Aboul-ela, 2004.
\newblock {Rational Design of Inhibitors of HIV-1 TAR RNA through the
  Stabilisation of Electrostatic "Hot Spots"}.
\newblock \emph{J. Mol. Biol.} 336:343--356.

\bibitem[Brion and Westhof(1997)]{WesthofARRBS97}
Brion, P., and E.~Westhof, 1997.
\newblock Hierarchy and dynamics of {RNA} folding.
\newblock \emph{Ann. Rev. Biophys. Biomol. Struct.} 26:113--137.

\bibitem[Brodsky et~al.(1998)Brodsky, Erlacher, and
  Williamson]{WilliamsonNAR98}
Brodsky, A.~S., H.~A. Erlacher, and J.~R. Williamson, 1998.
\newblock {NMR evidence for a base triple in the HIV-2 TAR C-G.C+ mutant-
  argininamide complex}.
\newblock \emph{Nucl. Acids Res.} 26:1991--1995.

\bibitem[Mathews et~al.(1999)Mathews, Sabina, Zuker, and Turner]{MathewsJMB99}
Mathews, D., J.~Sabina, M.~Zuker, and D.~Turner, 1999.
\newblock {Expanded Sequence Dependence of Thermodynamic Parameters Improves
  Prediction of RNA Secondary Structure}.
\newblock \emph{J. Mol. Biol.} 288:911--940.

\bibitem[Klimov and Thirumalai(2001)]{Klimov}
Klimov, D.~K., and D.~Thirumalai, 2001.
\newblock Lattice Model studies of force-induced unfolding of protein.
\newblock \emph{J. Phys. Chem. B} 105:6648--6654.

\bibitem[Lukavsky et~al.(2003)Lukavsky, Kim, Otto, and Puglisi]{PuglisiNSB03}
Lukavsky, P.~J., I.~Kim, G.~A. Otto, and J.~D. Puglisi, 2003.
\newblock {Structure of HCV IRES domain II determined by NMR}.
\newblock \emph{Nature. Struct. Biol.} 10:1033--1038.

\bibitem[Simpson et~al.(2000)Simpson, Tao, Leiman, Badasso, He, Jardine, Olson,
  Morais, Grimes, Anderson, Baker, and Rossmann]{RossmanNature00}
Simpson, A.~A., Y.~Tao, P.~G. Leiman, M.~O. Badasso, Y.~He, P.~J. Jardine,
  N.~H. Olson, M.~C. Morais, S.~Grimes, D.~L. Anderson, T.~S. Baker, and M.~G.
  Rossmann, 2000.
\newblock {Structure of the bacteriophage $\phi$29 DNA packaging motor}.
\newblock \emph{Nature} 408:745--750.

\bibitem[Rief et~al.(1997)Rief, Gautel, Oesterhelt, Fernandez, and
  Gaub]{GaubSCI97}
Rief, M., M.~Gautel, F.~Oesterhelt, J.~M. Fernandez, and H.~E. Gaub, 1997.
\newblock {Reversible Unfolding of Individual Titin Immunoglobulin Domains by
  AFM}.
\newblock \emph{Science} 276:1109--1111.

\bibitem[Cecconi et~al.(2005)Cecconi, Shank, Bustamante, and
  Marqusee]{MarquseeScience05}
Cecconi, C., E.~A. Shank, C.~Bustamante, and S.~Marqusee, 2005.
\newblock {Direct Observation of Three-State Folding of a Single Protein
  Molecule}.
\newblock \emph{Science} 309:2057--2060.

\end{thebibliography}
\end{document}